\documentclass{ws-ijmpa}
\usepackage{amsmath}
\usepackage{amssymb}
\usepackage{cite}
\begin{document}

\markboth{\footnotesize B.F.L. WARD}
{QUANTUM CORRECTIONS TO NEWTON'S LAW IN RESUMMED QUANTUM GRAVITY}

%
\catchline{}{}{}{}{}
%

\title{QUANTUM CORRECTIONS TO NEWTON'S LAW IN RESUMMED QUANTUM GRAVITY
}

\author{\footnotesize B.F.L. WARD}

\address{Department of Physics, Baylor University, One Bear Place \#97316\\
Waco, Texas 76798-7316,
USA
}

\maketitle


\begin{abstract}
We present the elements of resummed quantum gravity, a new
approach to quantum gravity based on the work of Feynman using the simplest
example of a scalar field as the representative matter.
We show that we get a UV finite quantum correction to Newton's law.

\end{abstract}

\section{Introduction}
Newton's law, one of the most basic laws in
physics, is a special case of the solutions of the
classical field equations of Albert Einstein's general theory of relativity.
Successful tests of Einstein's theory in classical physics
are described in Refs.~\cite{mtw,sw1,abs}.
Quantum mechanics, as formulated by Heisenberg and Schroedinger, following Bohr,has explained, in 
the Standard Model(SM)~\cite{sm}, 
all empirically established quantum phenomena except the quantum treatment of Newton's law. This obtains
even with the tremendous progress in quantum field theory,
superstrings~\cite{gsw,jp}, loop quantum gravity~\cite{lqg}, etc.
In this paper, we address this issue by using
a new approach~\cite{bw1,bw2} to quantum gravity(QG), 
building on previous work by Feynman~\cite{f1,f2}, to get a minimal
union of Bohr's and Einstein's ideas. 

From the view of the generic approaches~\cite{wein1}
to the attendant bad UV behavior of QG, 
our approach, based on YFS methods~\cite{yfs,yfs1},
is a new version of the resummation approach~\cite{wein1} and
allows us to make contact
with both the extended theory~\cite{wein1} 
and the asymptotic safety~\cite{wein1,laut,reuter2}
approaches and to address~\cite{bw1,bw2} issues in black hole physics, some of which relate to Hawking~\cite{hawk} radiation.\par

The description of our new resummed QG theory is already presented
in these {\it Proceedings} in Ref.~\cite{bw2}, to which we refer
the reader. Here, we go directly to the issue of the
quantum corrections to Newton's law.

\section{Quantum Corrections to Newton's Law in Resummed QG}

\begin{figure}
\begin{center}
\epsfig{file=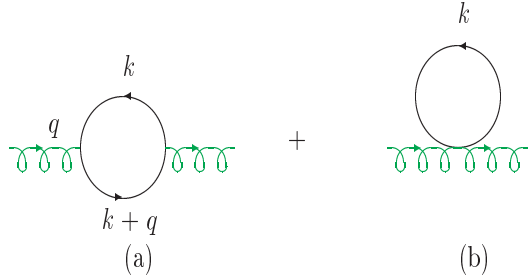,width=77mm,height=38mm}
\end{center}
\caption{\baselineskip=7mm     The scalar one-loop contribution to the
graviton propagator. $q$ is the 4-momentum of the graviton.}
\label{fig1}
\end{figure}

The model which we use~\cite{bw1,bw2} is gravity coupled to a 
scalar field as formulated in Refs.~\cite{f1,f2}. 
In our resummed QG theory, the graphs in Fig.~\ref{fig1}
\noindent 
become finite as we explain in Refs.~\cite{bw1,bw2}.
In this way, we get a UV finite quantum correction to Newton's law
{\em without modifying Einstein's theory}. In Refs.~\cite{bw1},
we show that this UV finiteness holds for all orders in the loop expansion.\par

Specifically, introducing the YFS resummed propagators as derived
in Refs.~\cite{bw2,bw1} into Fig. 1 yields
, by the standard methods~\cite{bw1}, 
that the graviton propagator denominator,
$q^2 +\frac{1}{2}q^4\Sigma^{T(2)}+i\epsilon$, 
is specified by
\begin{equation} 
-\frac{1}{2}\Sigma^{T(2)} \cong \frac{c_2}{360\pi M_{Pl}^2}
\label{sigma}
\end{equation}
for
$c_2 = \int^{\infty}_{0}dx x^3(1+x)^{-4-\lambda_c x}\cong 72.1$ 
where $\lambda_c=\frac{2m^2}{\pi M_{Pl}^2}$.
This implies the Newton potential
\begin{equation}
\Phi_{N}(r)= -\frac{G_N M_1M_2}{r}(1-e^{-ar})
\label{newtn}
\end{equation}
where $a=1/\sqrt{-\frac{1}{2}\Sigma^{T(2)}}\simeq 3.96 M_{Pl}$
when for definiteness we set $m\cong 120$GeV~\cite{lewwg}.
We note that
$c_2 \cong \ln\frac{1}{\lambda_c}-\ln\ln\frac{1}{\lambda_c}-\frac{\ln\ln\frac{1}{\lambda_c}}{\ln\frac{1}{\lambda_c}-\ln\ln\frac{1}{\lambda_c}}-\frac{11}{6}$.
Without resummation, $\lambda_c=0$,
and $c_2$ is infinite
and, as this is
the coefficient of $q^4$ in the inverse propagator, 
{\bf no renormalization of the field and/or of the mass could remove
such an infinity}. In our new approach, 
this infinity is absent.\par

We can make a 
cross check of our gauge invariant~\cite{bw1} analysis 
with the gauge invariant analysis of
Ref.~\cite{thvelt1} where the complete
result of the one-loop divergences of our scalar field coupled
to Einstein's gravity have been computed. Since $c_2$ diverges
without our resummation, it follows~\cite{bw1} that
we need to make the correspondence
between the poles in $n$, the dimension of space-time,
at $n=4$ calculated in Ref.~\cite{thvelt1}
and the leading 
log $\ln\frac{1}{\lambda_c}$. This implies~\cite{bw1}
\begin{equation}
\frac{1}{(2-n/2)} \leftrightarrow c_2,
\label{crsp1}
\end{equation}
so that, if we look at the limit $q^2\rightarrow 0$,
we find that the coefficient of $q^4$ 
in the graviton propagator denominator above
is $3/(2-n/2)$ times the coefficient 
of $c_2$ on the right-hand side of (\ref{sigma}), in
complete agreement with the result that is implied by
eq.(3.40) in Ref.~\cite{thvelt1}
, for example. 


Sub-Planck scale physics is accessible to point particle field theory
so that current superstring theories may be
phenomenological models
for a more fundamental theory (TUT=The Ultimate Theory) just as the old string theory~\cite{schw1} is such a model for QCD. Other types of correspondences
are not excluded here~\cite{superichep02}.
Our deep Euclidean studies are complementary 
to the low energy studies
of Ref.~\cite{dono1}. The effective cut-off which we generate dynamically
is at $M_{Pl}$ so that, while renormalizable quantum field theory (QFT)  
below $M_{Pl}$
is unaffected, some non-renormalizable QFT's are given new 
life here -- other problems notwithstanding.

Some phenomenological implications of (\ref{newtn}) are presented in Ref.~\cite{bw1,bw2}. To sum up, it appears that we may have indeed realized a minimal union
of the ideas of Bohr and Einstein.\par

\section*{Acknowledgments}
We thank Prof. S. Jadach for useful discussions.
This work was partly supported by the US Department of Energy Contract  
DE-FG05-91ER40627
and by NATO Grants PST.CLG.977751,980342.

\end{document}